# On Vulnerabilities, Constraints and Assumptions

Anil Bazaz and James D. Arthur
{abazaz, arthur}"at" vt.edu
Department of Computer Science
Virginia Tech

**Abstract**

*This report presents a taxonomy of vulnerabilities created as a part of an effort to develop a framework for deriving verification and validation strategies to assess software security. This taxonomy is grounded in a theoretical model of computing, which establishes the relationship between vulnerabilities, software applications and the computer system resources. This relationship illustrates that a software application is exploited by violating constraints imposed by computer system resources and assumptions made about their usage. In other words, a vulnerability exists in the software application if it allows violation of these constraints and assumptions. The taxonomy classifies these constraints and assumptions. The model also serves as a basis for the classification scheme the taxonomy uses, in which the computer system resources such as, memory, input/output, and cryptographic resources serve as categories and subcategories. Vulnerabilities, which are expressed in the form of constraints and assumptions, are classified according to these categories and subcategories. This taxonomy is both novel and distinctively different from other taxonomies found in the literature.*

## 1. Introduction

In recent years, there have been numerous reported exploits targeting software applications[10, 18]. These applications are exploited by using vulnerabilities present in them. A vulnerability is defined as a state of the system, but what differentiates this state from any other state is the fact that it is possible to jump to an incorrect system state from it [9]. In other words, a vulnerability is a defect which, when exercised, can produce undesired and incorrect behavior [23]. It is important to note the difference between a vulnerability and an exploit. An exploit consists of a vulnerability present in the software application and a method that is used to exploit this vulnerability. Thus, an exploit occurs when we actually apply the method to execute the vulnerability present in the software application.

Since a vulnerability is fundamental to exploiting a software application, we can prevent an exploit if we are able to identify and subsequently eliminate vulnerabilities present in a software application. However, identifying vulnerabilities is a very difficult task because of a number of contributing factors. The primary ones are:

1. *Complexity of software applications:* Modern software applications are large and complex and have hundreds of thousands of lines of code. Furthermore, their complexity increases as they use services provided by other applications, making it very difficult to find a vulnerability.

2. *Number of vulnerabilities:* Because there are numerous known vulnerabilities, it is not practical simply to make a list of all of them and use it to identify the ones present in a software application.

3. *Complexity of vulnerabilities:* Some vulnerabilities, such as time of check time of use flaw, are so complicated in nature that they are difficult to identify, because they involve multiple software components interacting together to produce the vulnerable system state.

All of these factors make identifying vulnerabilities a formidable task. However, we can facilitate this identification by grouping vulnerabilities into categories. This allows the development of a systematic approach to testing a software application for the presence of vulnerabilities. But simply grouping vulnerabilities together does not help in identifying vulnerabilities present in a software application. In order to simplify the identification of vulnerabilities, we need to acquire an in-depth understanding of vulnerabilities that includes how they are exploited and their relationship with software applications and computer system resources. This understanding provides a new perspective on vulnerabilities, which, in turn, simplifies identification of vulnerabilities present in a software application.

In this report, we present a taxonomy of vulnerabilities that provides an intelligent classification of vulnerabilities. The taxonomy is grounded in a theoretical model of computing [5], which establishes the relationship between vulnerabilities, software applications and computer system resources. This relationship illustrates that software applications are exploited by violating constraints imposed by computer system resources and assumptions made about the usage of these resources. These constraints and assumptions form the reason that a software application is vulnerable to exploits. In other words, a vulnerability exists in the software application if it allows violation of a constraint or an assumption. This taxonomy classifies these constraints and assumptions. The theoretical model also provides the classification scheme used by our taxonomy, which consists of using computer system resources as categories. The three top level categories of our taxonomy are: (1) main memory, (2) Input/Output (I/O), and (3) cryptographic resources. Each of these categories is further divided into subcategories in which the constraints and assumptions are finally classified.

We have developed this taxonomy to assist in the development of a framework to derive verification and validation (V&V) strategies to assess software security [3]. This taxonomy is the most important part of the framework, because it structures vulnerabilities into categories, thereby enabling us to develop a systematic approach for assessing software security. Furthermore, the taxonomy simplifies the testing of a software application for the presence of vulnerabilities. Since the vulnerabilities listed in the taxonomy are in the form of violable constraints and assumptions, we can simply test if the software application allows violation of a constraint or an assumption, and a vulnerability exists if either can be violated.

The rest of this report is organized as follows: Section 1.1 establishes the level of abstraction of the taxonomy; Section 1.2 presents the theoretical model; Section 2 provides a description of some existing taxonomies as related work; Section 3 details the taxonomy; and Section 4 presents conclusions and future work.

**1.1 Level of abstraction of the taxonomy**

Since the taxonomy applies to the software application, it is first necessary to define the software application, which we do in terms of its relationship with the other components of the computer system. Modern computer systems can be visualized as a hierarchy of components. Figure 1.1 presents a rendering of this hierarchy as a layered pyramid. At the bottom is the hardware or physical devices, such as memory, hard disk, processor, buses, and so forth. The operating system, which manages these hardware devices and provides an interface to use them, occupies the layer above hardware. Software applications occupy the layer above the operating system and provide specific services to the computer system users and to other software applications. These applications are comprised of one or more software processes. The term software process, as used here, refers to an executing program and is central to our taxonomy. This visualization of the software process defines the level of abstraction of the taxonomy. The taxonomy applies to any software process that executes above the level of the operating system. The process does utilize the hardware resources, but the underlying operating system defines its view of these resources.

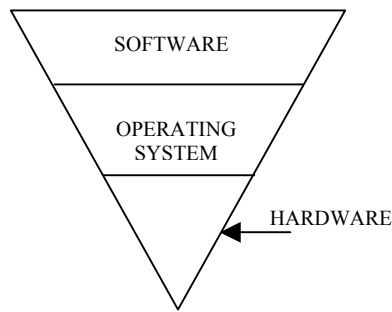

**Figure 1.1 Hierarchal view of computer system**

## 1.2 Theoretical Model of Computing

In *Modeling Security Vulnerabilities: A Constraints and Assumptions Perspective* [5], we introduced a theoretical model of computing, which enables us to characterize vulnerabilities as constraints and assumptions. This model takes a high level view of a software process executing on the computer system (Figure 1.2). The process can be visualized as a black box entity that takes input, performs some operations, and produces an output. However, it does not execute in vacuum. To perform its requisite operations, it requires resources such as *memory* to store data, instructions and other execution specific parameters and *I/O* to receive and store input, store output, present output etc., as well as *cryptographic resources*[1] to store secrets, ensure data integrity and so forth. A software process uses one or more of these to carry out its requisite functions.

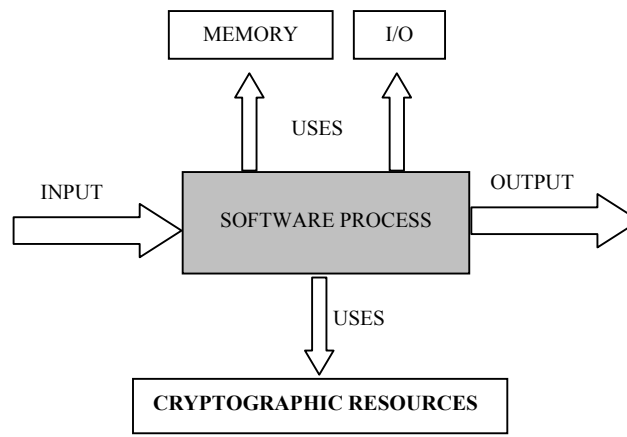

**Figure 1.2 High level view of software process executing on a computer system**

A software process cannot, however, use any of these resources unrestrained. The resources have to be used within certain rules and limits defined by the resources themselves or by the operating system that provides an interface to these resources. These rules and limits manifest themselves in the form of constraints that a software process has to adhere to in order to function correctly. In addition to these constraints, a software process also makes assumptions about the resources it is utilizing and has to ensure that the assumptions it is making are correct and do hold under all circumstances. It is our contention that a vulnerable state arises when a software process fails to enforce the constraints imposed by the resources and/or makes violable assumptions about their usage. In other words, a vulnerability exists if the software process allows violation of these constraints and assumptions.

To further illustrate the concept, let us consider the example of the buffer overflow. The system state can be described as a buffer that is allocated on the program stack with no bounds checking being performed on it. The program stack (part of the memory) imposes the following constraint on the software process: d*ata accepted as input by the process and assigned to a buffer will occupy and modify only specific locations allocated to the buffer.* Neither the hardware nor the operating system enforce this constraint, even though the software process is expected to enforce and adhere to it. However, because the software process is using an unbounded buffer, the constraint can be easily violated. The violable constraint is the reason that this system state is vulnerable, which leads to the undesired and incorrect functionality when the process accepts and stores data that is larger than the buffer, thereby overwriting data that lies beyond the bounds of the buffer.

---

[1] Please note that cryptographic algorithms and protocols are considered as resources. It is our contention that these algorithms and protocols should be treated as a resource and should not, under any circumstances, be developed by software developers.

## 2. Related Work

In 1976 Abbot et al. [1] developed one of the first security taxonomies as a part of the RISOS project. It categorizes operating system integrity flaws into seven categories. This taxonomy represents one of the first steps in trying to understand and structure the field of software security. But Bishop [9] has shown that the taxonomy is ambiguous, because a single flaw can be categorized into multiple categories.

In 1995, Aslam [19] presented a taxonomy of security faults. This taxonomy, which Aslam developed to organize information being stored in a vulnerability database, consists of three top level categories: operational faults (configuration errors), coding faults, and environment faults. Although this is a good taxonomy, Bishop [9] shows that it too is ambiguous because it also allows a single vulnerability to be classified into multiple categories.

Krsul [13] extends Aslam's work by further classifying the environmental faults category. The top level categories of his taxonomy consist of environmental objects, each of which is defined as an entity that contains or receives information and has a unique name and a set of operations that can be carried on it. Associated with each object are its attributes, each of which is defined as the data component of the object. These attributes form the second level categories of the taxonomy and can be further refined if required. Each of these attributes has associated with it assumptions the programmers make, which Krsul asserts are responsible for environmental vulnerabilities.

Although the most detailed and complicated of those included in this section, Krsul's taxonomy has some shortcomings. First, the definition of objects and their attributes leaves too much leeway for interpretation, because whether a particular entity is to be considered as an object or an attribute depends on interpretation. For example, an environment variable can be considered as an attribute of the object running program or it can be considered as an object by itself. This introduces confusion and ambiguity into the taxonomy. Furthermore, the taxonomy lacks a theoretical foundation that ties it together and defines the relationship between assumptions, attributes and objects. However, despite its shortcomings and complications, Krsul's taxonomy is one of the best developed to date, because it provides us with a significant insight into vulnerabilities by focusing on assumptions responsible for them.

Other notable taxonomies developed to date include Newman and Parker's [16] taxonomy of computer misuse techniques, Lindqvist and Jonsson's [14] extension of Newman and Parker's taxonomy, VERDICT [15], and so forth. Although, these contain interesting insights, we restrict our discussion because of space limitations.

The taxonomy presented in this report is distinctively different from those developed to date. It is grounded in a theoretical foundation that underlies the taxonomy. Moreover, the theoretical foundation provides us with an in-depth understanding of vulnerabilities and serves as a basis for the novel classification scheme our taxonomy uses. Another feature that distinguishes our taxonomy from any other is that it classifies vulnerabilities by characterizing them as constraints and assumptions. Additionally, we define the level of abstraction of the taxonomy. It applies to any software application executing above the level of the operating system. Furthermore, our taxonomy is not ambiguous, because a single constraint or assumption can be classified only in a single category, which is a direct result of using resources to classify constraints and assumptions. Since constraints are imposed by resources and assumptions are made about their usage, a single constraint or assumption can be associated only with a single resource, which implies that it can only be classified in a single category.

**3.0 Taxonomy**

This section presents our taxonomy of vulnerabilities. This taxonomy applies to any software application executing above the level of the operating system. The theoretical model, outlined in Section 1.2, lays the foundation for the taxonomy by establishing a relationship between vulnerabilities and system resources. This relationship enables us to characterize vulnerabilities as constraints imposed by computer system resources and assumptions made about the usage of these resources. Our taxonomy targets these constraints and assumptions. In other words, vulnerabilities are reduced to constraints and assumptions, which the taxonomy then classifies. We identify the constraints and assumptions currently classified by the taxonomy by analyzing vulnerabilities, which, in turn, have been identified by analyzing security exploits. We obtained the exploits from various sources including, but not limited to, books [12, 22], mailing lists [10], websites [18, 17], and so forth.

Figure 3.1 illustrates the taxonomy of vulnerabilities. The theoretical model shows that constraints and assumptions are bound to the resources, an insight we use to derive the classification scheme of the taxonomy. This scheme consists of using the resources as categories of the taxonomy. Therefore, the three top-level categories of our taxonomy are (1) Main Memory, (2) I/O, and (3) Cryptographic Resources. Each category of the taxonomy is divided into subcategories, which themselves are resources, but each represents a different component of the top level category. The high level of abstraction of the top level categories, as well as their various components, makes this decomposition necessary. Each of the components has its own set of characteristics and the software process utilizes each for a different purpose. Consequently, each component has different constraints and assumptions associated with it, which necessitate different subcategories for different components.

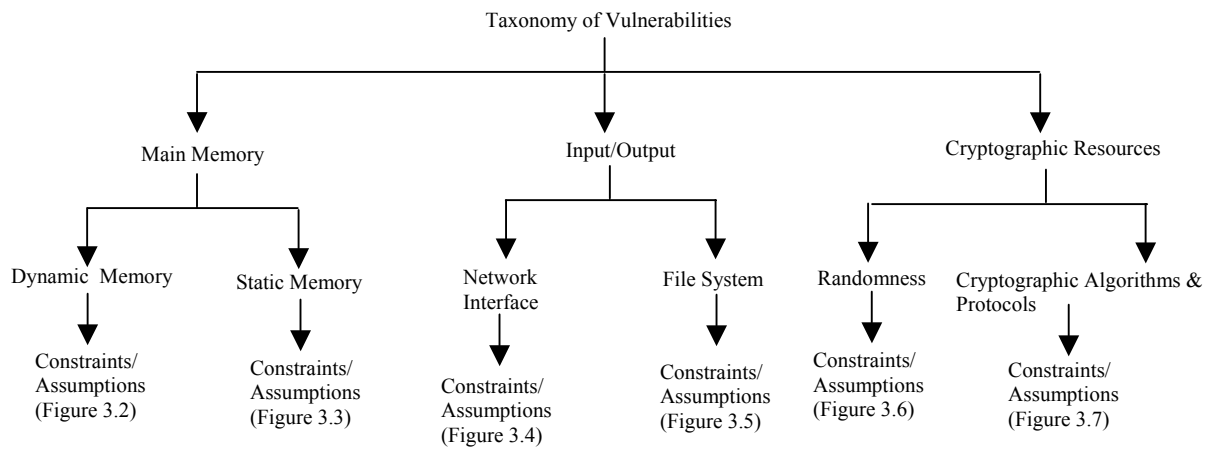

**Figure 3.1 Taxonomy of vulnerabilities.**

Our taxonomy classifies a constraint or an assumption into one of the subcategories, based on the resource with which it is associated. For example, an assumption such as *the software process will be provided with the required dynamic memory necessary for its execution* is categorized under the dynamic memory subcategory, as the resource in question is the dynamic memory.

It is important to note that our taxonomy views the resources from the same point of view as the software process. That is, the taxonomy is not concerned with the physical view or the operating system view of the resources, because the software process is using the resources and both constraints and assumptions apply to this software process.

The rest of this section describes each category, its subcategories, and lists constraints and assumptions associated with each subcategory. Additionally, because constraints and assumptions are described using system terminology, a short explanation follows each constraint and assumption. In order to illustrate the relation between constraints and assumptions and vulnerabilities, we also provide an example a vulnerability resulting from the violation of a constraint or an assumption for each category.

**3.1 Main Memory**

The main memory refers to the storage space the software process uses, while in execution, to store input, generated output, execution specific parameters, instructions, and other objects necessary for the execution of the process. A modern computer runs a number of processes at any given time. Each of these processes is allocated its own share of memory, which is referred to as virtual memory. Hence, each process has its own view of memory that is independent of the physical view. The underlying operating system supplies it with this view. The taxonomy takes this view of memory when it refers to main memory as a category.

As a category of this taxonomy, the main memory is divided into two subcategories, dynamic memory and static memory. Subsections 3.1.1 and 3.1.2 present the subcategories and the constraints and assumptions associated with them. Subsection 3.1.3 presents an example of a vulnerability resulting from the violation of a constraint associated with dynamic memory subcategory.

**3.1.1 Dynamic Memory**

Dynamic memory, the first subcategory of main memory, is, as the name suggests, the memory component whose size changes as the process executes. It consists of two subcomponents, the program stack and the heap, both of which store variables while the process is in execution. The program stack or execution stack is a contiguous block of memory used by the operating system and the process to store process data, such as local variables, function frames, return addresses, environment variables, program name, and so forth. The heap is a block of memory used to store dynamically allocated variables. For example, in C language the heap stores variables that are allocated using *malloc( )* call.

Figure 3.2 presents the constraints and assumptions associated with dynamic memory. Most of these constraints and assumptions are applicable to data held by variables stored on the dynamic memory. They are simple and self-explanatory. However failure to take each into account gives rise to a vulnerable condition.

# Dynamic Memory

1. ***Data accepted as input by the process and assigned to a buffer will occupy and modify only specific locations allocated to the buffer.***
   This assumption is violated if the software process allows data larger than the size of the buffer to be written to the buffer. Since the data is larger than the buffer, it will overwrite memory that lies beyond the bounds of the buffer.

2. ***The process will not interpret data present on the dynamic memory as executable code.***
   This assumption is violated if the process is made to interpret data present on the dynamic memory as executable code, which, in turn, can be accomplished by changing process variables it holds, such as, return addresses, exception pointers, and so on.

3. ***Environment variables being used by the process have expected format and values.***
   This is a violable assumption, because a hostile entity can easily change the environment variables before the process begins execution. These variables are provided by the operating system and define the behavior of the process. A violation occurs when the process uses these variables and makes assumptions regarding their format and values.

4. ***The process will be provided with the dynamic memory that it requests.***
   This is a violable assumption, because the amount of dynamic memory available to a process is limited and depends on the total amount of memory available on the computer system and the number of running processes. A violation occurs when the process assumes that it has access to an unlimited amount of memory.

5. ***Data present on the dynamic memory cannot be observed while the process is in execution.***
   This is a violable assumption, because a hostile entity can run the process in a controlled environment and observe the contents of the dynamic memory, including any privileged data[3] it holds.

6. ***Data owned by the process and stored on the dynamic memory cannot be accessed after the process frees the memory.***
   This is a violable assumption, because the memory being used by the process is not erased after the process frees it. Hence, another process, if allocated the same physical memory, can access the data left over by the previous process.

7. ***A pointer variable being used by the process references a legal memory location.***
   This is a violable assumption, because a pointer variable can point to any memory location, including memory locations outside of the process address space. Furthermore, it can reference wrong variables, thereby creating illegal memory references.

8. ***A memory pointer returned by the underlying operating system does not point to zero bytes of memory.***
   This is a violable assumption, because the operating system can provide the process with a pointer that points to zero bytes of memory. Using this pointer will cause illegal memory references or overwriting of memory locations being used by other process variables.

9. ***A pointer variable being used by the process cannot reference itself.***
   This assumption is violated if a pointer variable references itself. The consequences of this violation vary from garbage value being written to the memory location to the process going into an infinite loop.

10. ***Data accepted by the process will not be interpreted as a format string by the I/O routines.***
    This constraint is violated if a process accepts input and interprets it as a format string. A violation will at the very least reveal the contents of the process stack. Additionally, a hostile entity can provide the process a specially crafted format string that allows it to write data to the process stack.

**Dynamic Memory (Cont.)**

11. ***The value of an integer variable/expression (signed & unsigned) accepted/calculated by the process cannot be greater (less) than the maximum (minimum) value that can be stored in the integer variable.***
    This is a violable assumption, because the maximum (minimum) value of an integer variable is determined by the amount of storage space provided to it by the underlying operating system. If the process tries to store a value that requires more storage space than is allocated to an integer, then the higher order bits of this value are dropped, resulting in a wrong value being stored.

12. ***An integer variable/expression used by the process as the index to a buffer will only hold values that allow it access to the memory locations assigned to the buffer.***
    This constraint is violated if the process does not restrict the value of the integer variable being used as an index to the array. A hostile entity can provide the process with any integral value to use as an index, which, in turn, gives it access to any memory location that lies beyond the bounds of the array.

13. ***An integer variable/expression used by the process to indicate length/quantity of any object will not hold negative values.***
    This constraint is violated if the process uses an integer value to indicate length/quantity of an object and does not restrict the value to only positive values. A hostile entity can use a negative value to indicate the length of the object, thereby creating an error condition.

**Figure 3.2 Constraints/Assumptions associated with dynamic memory**

**3.1.2 Static Memory**

Static memory, the second subcategory of the main memory, is, as the name suggests, static in size. In other words, the size of the memory is fixed before the process begins execution and does not change as the execution proceeds. There are two major components of static memory, the data segment and the block storage segment (BSS). The data segment stores initialized global variables and the BSS stores un-initialized global variables. The software process uses the two components in a similar way, to store global data whose size is fixed before execution starts and does not change during execution of the process. Figure 3.3 presents the constraints and assumptions associated with static memory. Again, the constraints and assumptions are simple, but failure to take them into account gives rise to potential vulnerable states.

**Static Memory**

1. ***Data accepted as input by the process and assigned to a buffer occupies and modifies only specific locations allocated to buffer on the static memory.***

    This assumption is violated if the software process allows data larger than the size of the buffer to be written to the buffer. Because the data is larger than the buffer, it will overwrite memory that lies beyond the bounds of the buffer.

2. ***Data held on the static memory cannot be observed while the process is in execution.***

    This is a violable assumption, because a hostile entity can run the process in a controlled environment and observe the contents of the static memory, thereby gaining access to any privileged data held in static memory.

**Figure 3.3 Constraints/Assumptions associated with static memory**

### 3.1.3 An Example of a Main Memory Vulnerability

Item 12 in figure 3.2 is used here to provide a detailed example of a vulnerability resulting from a violable constraint associated with dynamic memory. The vulnerable state occurs when a software process accepts from users two integer values, a position to store data and the value of the data itself. The software process stores the data in an integer array allocated on the program stack and uses the position value as an index to the array. Because it does not restrict the values of either the position or the data, users are free to enter any integral value they desire.

This vulnerability results from a software process allowing any value as the index of the array. Users can exploit this vulnerability by providing the software process with an index value greater than the size of the array, which permits them to write to any memory location beyond the array. Because users also provide data corresponding to an index, they can write any value to any memory location. Thus, the users can overwrite the return address on the program stack to the value of their choice, which gives them the ability to execute arbitrary code with the privileges of the software process. In other words, the software process creates a vulnerable state by failing to ensure that the constraint is not violated.

### 3.2 Input/Output (I/O)

The second category of the taxonomy is I/O. The software process uses I/O resources to store input and output, present output, receive input, and to communicate output. I/O, as a category of this taxonomy, is divided into two subcategories, filesystem and network interface. Subsections 3.2.1 and 3.2.2 present the subcategories and the constraints and assumptions associated with them. Subsection 3.2.3 presents an example of violation of an assumption associated with the filesystem subcategory.

### 3.2.1 Filesystem

The filesytem, the first subcategory of I/O, refers to the encoding of the addressing scheme, boot block, super block, inode structure, storage blocks, and so forth. Different filesystem types use different encoding schemes. For example, ext3, FAT, and NTFS all use different encoding schemes. However, a software process, while using a filesystem as a resource, does not perceive it as an encoding scheme; instead it views the filesystem as a hierarchy of connected directories containing metadata about files and files themselves. The taxonomy adopts this view when referring to the filesystem as a subcategory.

## Filesystem

1. *Access permissions assigned to newly created files/directories are such that only the required principals have access to them.*
   This constraint is violated if a software process creates new files and directories without assigning them proper access permissions, which, in turn, allows a hostile entity to access them. This results in an error condition when a hostile entity actually reads or modifies these files and directories.

2. *Access permissions of the files/directories being used by the process are such that only the required principals have access to them.*
   This is a violable constraint, because if a software process uses file/directories, already present on the filesystem, which do not have proper access permissions, then there exists a probability that a hostile entity has already read or modified these files. A violation occurs when the entity actually reads or modifies these files.

3. *A file being created by the process does not have the same name as an already existing file.*
   This is a violable assumption, because a hostile entity can create a file with the same name as a file being created by the software process and place it in the same directory where the software process was going to place its file. Depending on the underlying operating system, the consequences of this attack vary from the process using the file placed by the hostile entity to the process terminating execution.

4. *A filename (including path) being used by the process is not a link that points to another file for which the user, executing the process, does not have the required access permissions.*
   This is a violable assumption, because a hostile entity can provide the software process with a file that is a link to another file. If the process has more privileges than the entity, then the entity can point the link to a file to which it does not have access, thereby resulting in the entity gaining improper access to the file pointed to by the link.

5. *Files created/populated by a principal other than the process and being used by the process will have expected format and data.*
   This is a violable assumption, because a hostile entity can provide the software process with specially crafted files containing corrupt data. A violation occurs when the software process uses these corrupt files.

6. *Files being used by the process cannot be observed/ modified/replaced while the process is in execution.*
   This constraint is violated if the software process does not lock the files that it is using. A violation occurs when a hostile entity can read, modify or replace these files while the process is in execution.

7. *Files/directories being used by the process and stored on the file-system (information used by the process over multiple runs) cannot be observed/modified/replaced before the process starts execution.*
   This is a violable assumption, because a hostile entity can observe, modify or replace files created by the software process and stored on the filesystem after the process stops execution and before the next run.

8. *Data held by files owned/used by the process cannot be accessed after the process deletes them.*
   This is a violable constraint, because files stored on the filesystem are not erased after the software process deletes them. The operating system simply deletes their name from the list of existing files, but anyone with proper access permissions can access this data by directly accessing the physical storage.

9. *The process will be provided with the file-system space that it requests.*
   This is a violable constraint, because filesystem space is a limited resource and the process will not always be provided with the filesystem space that it requires.

10. *Files having proprietary or obscure file formats cannot be understood and modified.*
    This is a violable constraint, because proprietary or obscure file formats are not enough to keep the contents of a file secret. Tools and techniques a hostile entity can use to reverse engineer these formats to reveal the contents of the files exist.

**Figure 3.4 Constraints/Assumptions associated with filesystem**

Figure 3.4 presents the constraints and assumptions associated with the filesystem. Most of the constraints and assumptions associated with the filesystem are simple and directly involve files and directories. Historically, software processes have been particularly vulnerable to exploits involving files/directories, because the filesystem itself does not implicitly enforce a number of constraints. Furthermore, by failing to take these constraints into account and making assumptions regarding filesystem usage, the software process creates states wherein these constraints and assumptions can be violated with relative ease. This, of course, creates vulnerabilities and, in turn, permits exploits.

### 3.2.2 Network Interface

The network interface, the second subcategory of I/O, refers to the interface used by the software process to send and receive data. This interface is in the form of ports. A port is an end of a logical connection, which a software process uses to communicate over the network. Ports are numbered from 1 to 65535, with different ports being used by different applications for different services. A software process can attach to any open port and use it to send and receive data. Therefore, from the perspective of the software process, the network interface is simply a resource that can be used to send and receive data. The taxonomy takes this perspective when referring to network interface as a subcategory.

**Network Interface**

1. ***The data received by the software process through the network interface can be trusted to be unread and unmodified.***
   This is a violable assumption, because a hostile entity can intercept and read or modify data that has been sent to the software process by sitting between the sender and the software process. A violation occurs when a hostile entity reads or modifies data that has been sent to the software process.

2. ***The data received by the software process through the network interface is from a legitimate client or peer or server and has expected format and length.***
   This is a violable assumption, because any entity, even a hostile one, can send data using the network interface if it knows the IP address of the host machine and the port number being used by the software process. A violation occurs when a hostile entity sends corrupt data to the software process.

3. ***The data sent by the software process via the network interface will not be read/modified before it reaches its destination.***
   This is a violable assumption, because a hostile entity can intercept and read or modify data being sent by the software process by sitting between the software process and receiver of data. A violation occurs when a hostile entity reads or modifies data being sent by the software process.

4. ***The software process will be able to utilize the network interface to send and receive data.***
   This is a violable constraint, because no guarantee that the software process will always be provided with access to the network interface exists. A violation occurs if the software process does not take this constraint into account and is not able to use the network interface.

**Network Interface**

5. ***The byte order of numerical data accepted from the network interface is same as that of the host machine.***
   This is a violable assumption, because the format of numerical data of a host machine and the network can be different and the software process has to change the format before using it. Making this assumption results in wrong numerical values being used by the software process.

**Figure 3.5 Constraints/Assumptions associated with network interface**

Figure 3.5 presents the constraints and assumptions associated with the network interface. Considering the large number of exploits that have taken place in the past, the total number of constraints and assumptions seems relatively small. This stems from the fact, however, that the network interface, the operating system, and the software process do not typically enforce these constraints and assumptions. The software process, without realizing that the constraints and assumptions can be easily violated, makes decisions based on the perception of "valid" data being received from the network interface, which, in turn, can lead to vulnerable system states.

### 3.2.3 An Example of an I/O Vulnerability

Item 4 in Figure 3.4 is used here to provide a detailed example of a vulnerability resulting from a violable assumption associated with the filesystem. The vulnerable system state occurs when a software process executes with setuid permissions. In other words, although a user is executing the process, it has effective root user privileges, which give the process access to all resources of the computer system. The software process, in course of its execution, asks the user to provide it with the name of a file to which it has to write data. However, before opening the file, the process fails to check the access permissions of the file pointed to by the symbolic link; it simply opens the file and writes data to it.

This vulnerability results from the software process not checking if the user has the required access permissions to the file pointed to by the symbolic link. A user can exploit this vulnerability by providing the software process with the name of a symbolic link, which points to a file to which the user does not have the required access permissions. If, for instance, the symbolic link points to the system password file, the software process will follow the link, open the system password file, and write data to it. Depending on the data the process is writing, the consequences of this exploit vary from trashing of system password file to a user gaining root privileges. Therefore, by not taking this assumption into account, the software process fails to ensure that it is not violated, which creates the vulnerability.

### 3.3 Cryptographic Resources

Cryptographic resources, the third category of our taxonomy, refers to the algorithms and protocols contributed by the discipline of cryptography. In computer systems, cryptography provides the capability for securing data and resources in the areas of confidentiality, authentication, integrity, and nonrepudiation [22]. Encryption algorithms are used to ensure confidentiality, which means that only authorized entities are able to understand data. Authentication amounts to ensuring that only authorized entities are able to access data/resources or to supply data. Algorithms, such as hashes, checksums and so forth, are used to ensure data integrity, which means only authorized entities should be able to modify data. Nonrepudiation, which concerns the communication of messages, means that the sender of a message should be able to prove that the receiver received the message; conversely the receiver should be able to prove that the sender actually sent the message. Nonrepudiation is very difficult to prove, given current cryptographic algorithms and protocols. However, digital signatures, PKI, and so forth, do solve the problem to a certain extent.

The two subcategories of the cryptographic resources category are randomness resources and cryptographic algorithms and protocols. Subsections 3.3.1 and 3.3.2 present the subcategories and the constraints and assumptions associated with them. Subsection 3.3.3 presents an example of a vulnerability resulting from the violation of an assumption associated with cryptographic algorithms and protocols subcategory.

### 3.3.1 Randomness resources

Randomness, the first subcategory of cryptographic resources, refers to the generation and use of random numbers, which are a series of numbers whose values are uniformly distributed over a set and where it is impossible to predict the next number in the series. True random numbers can be generated only by using physical phenomena like radioactive decays. Since using physical phenomena as a source of random numbers is not practical, algorithms have been developed for generating random numbers. These algorithms, called Pseudo-Random Number Generators (PRNGs), accept input, called a seed, and use it to generate a series of random numbers. The series generated is dependent on the seed provided to the PRNG. To generate an unpredictable series of numbers, an unpredictable seed should be used. The seed for PRNGs is generated using hard to guess events that occur in a computer system, such as press of a key on the keyboard, CPU scheduling, difference between CPU timer and interrupt timer, and so forth. The generation and use of random numbers is one of the least understood aspects in the design and implementation of a software system. Consequently, software processes tend to ignore the constraints associated with randomness and make assumptions that can be violated easily.

## Randomness Resources

1. ***The series of random data being produced by the PRNG is unpredictable (assuming unpredictable seed).***
   This is a violable assumption, because there exist PRNGs that produce predictable random data series. These PRNGs produce data that is random in the sense that each number has an equal probability of being the next number in the series, but it is computationally feasible to predict the next number in the series. Hence, the random data produced by these generators is predictable and cannot be used for cryptographic purposes.

2. ***The seed being used by the PRNG is unpredictable.***
   This assumption is violated if the seed being used by the PRNG is predictable. Random data being produced by the PRNG is dependent upon the seed provided as input to the PRNG. Thus, using an unpredictable seed is a key requirement for producing unpredictable random data series. A predictable seed would result in a predictable random data series, which cannot be used for cryptographic purposes.

3. ***The process will have easy access to entropic data on a computer system.***
   This is a violable assumption, because computers are deterministic machines. Therefore, it is very difficult to have to access entropic data on a computer system.

4. ***The process will be able to accurately estimate entropy of a data set.***
   This is a violable assumption, because currently available approaches for estimating entropy are very difficult to implement and provide only a coarse approximation of the value of entropy for a data set. Hence, a process, even while employing these approaches, should be conservative in estimating entropy, because estimating a higher value of entropy than is actually present leads to a false sense of security.

5. ***User selected passwords/keys will have a sufficient amount of entropy.***
   This is a violable assumption, because user selected passwords/keys do not have sufficient amount of entropy and typically are highly predictable. Hence, they should not be used in cryptographic algorithms or protocols that require highly entropic keys/passwords.

6. ***If two different seeds are provided to the PRNG, it is computationally infeasible to produce the same series of data both times.***
   This is a violable assumption because of the structure of the data series some PRNGs produce, which can be visualized as a series of numbers on the circumference of a circle. The seed, given as input to these PRNGs, selects a point on this circle from where the PRNG starts to output random numbers. It is possible, especially if the circle is small, for two different seeds to select the same point on the circle (modulo arithmetic) and produce the same random data series. These PRNGs are not suitable for cryptographic purposes and can be identified from current cryptographic literature.

7. ***Given that the PRNG is continuously producing random data, it is computationally infeasible to produce the same sequence of random data after some time.***
   This is a violable assumption, because it is possible that PRNGs, which adhere to the circular structure visualized in item 6 and produce random data continuously, will repeat the data series after some time.

**Figure 3.6 Constraints/Assumptions associated with randomness resources**

Figure 3.6 presents the constraints and assumptions associated with randomness resources. Because the taxonomy considers random numbers a resource, it is assumed that the software process is using standard random number generation techniques. Therefore, none of the constraints and assumptions associated with randomness resources addresses techniques for random number generation. This subcategory addresses the constraints inherent in the nature of random numbers and the assumptions made by the software process while using them.

## 3.3.2 Cryptographic algorithms and protocols

The second subcategory of cryptographic resources, cryptographic algorithms and protocols, refers to algorithms and protocols that provide to the software process the services of confidentiality, integrity, authentication and nonrepudiation. This subcategory includes encryption algorithms, such as DES, RC4 and RSA, authentication protocols, such as Kerberos, and cryptographic checksums and hashing algorithms, such as MD5 and SHA1. It also includes protocols and algorithms, such as digital signatures and PKI that promote nonrepudiation. This taxonomy treats cryptographic algorithms and protocols as resources, just as it does randomness.

### Cryptographic Algorithms and Protocols

1. ***Random data being used by the cryptographic algorithm/protocol is unpredictable.***
   This is a violable assumption, because there exist PRNGs that produce predictable random data series. Cryptographic algorithms and protocols require a random data series to exhibit two properties: (1) the data series is statistically random. That is, each number in the set of numbers has an equal probability of being the next number in the random data series, and (2) the next number in the series is unpredictable. There are PRNGs that produce random data that is statistically random but not unpredictable. These PRNGs cannot be used as a source of random data for cryptographic purposes.

2. ***The length of the key being used by the cryptographic algorithm and protocol is sufficient.***
   This assumption is violated if the minimum length of the key, being used for a cryptographic algorithm or protocol, is less than the current standard. Most cryptographic algorithms and protocols use keys for a variety of purposes, such as keeping secrets, restricting access and so on. The length of these keys is a critical factor in security of these algorithms or protocols. Typically, if a small key is used, the algorithm or protocol can be compromised easily. The length of the key required for keeping the algorithm or protocol secure changes with time and can be found in the current cryptographic literature. Using a key that is smaller than the current standard puts the algorithm and protocol at risk of being compromised.

3. ***The hashing algorithm will not produce same hash for two different inputs.***
   This is a violable assumption, because there is a possibility that a hashing algorithm produces same hash for two different input texts. These algorithms are considered as compromised and cannot be used for cryptographic purposes. Information about compromised hashing algorithms is available in current cryptographic literature.

4. ***The process cannot use encryption to ensure data integrity.***
   This is a violable constraint, because encryption only ensures data confidentiality by changing plain text into cipher text. No one can make sense of cipher text without changing it back to plain text. But anyone with access to cipher text can change it even if it does not make sense. Hence, encryption just ensures data confidentiality and not data integrity.

5. ***The process cannot use a key more than once for a stream cipher.***
   This is a violable constraint, because using a key more than once for a stream cipher can compromise the cipher text. Stream ciphers use a key to produce a stream of random data, which they xor with the plain text to produce cipher text. If the process uses the same key more than once, then the resultant random data series is same, which, in turn, implies that multiple instances of plain text data will be xored with the same random data series. There are known crypto-analytical techniques for compromising cipher texts produced by xoring multiple instances of plain texts with same random data series. Hence, a process should never use same key more than once for stream ciphers.

6. ***The process cannot use one time pads to encrypt a large quantity of data.***
   This is a violable constraint, because one time pads require truly random or close to truly random data that has bit length equal to that of the plain text. It is very difficult to have access to large quantity of high quality random data. Hence, one time pads cannot be used to encrypt large quantity of data.

7. ***The process cannot use keys that are self reported by a client or a server.***
   This is a violable constraint, because a hostile entity can masquerade as a client or a server and send its own keys to the process. Hence, any key that is self reported by a client or server cannot be trusted without some form of validation.

8. ***The process cannot use obfuscation instead of encryption to ensure confidentiality.***
   This is a violable constraint, because there exist tools and techniques that we can use to reverse engineer obfuscated data to reveal any privileged data. Obfuscation is the process of changing data so as to make it difficult to perceive or understand. In this way, it can be visualized as a much weaker form of encryption, which can be compromised easily. Thus, a process should not use obfuscation to ensure data confidentiality.

9. ***The process cannot store keys/passwords in clear text.***
   This constraint is violated if a process stores keys/passwords in plain text. Keys/ passwords are very important for the security of cryptographic algorithms and protocols. Compromise of a key or a password essentially means complete compromise of the cryptographic algorithm or protocol. Hence, they should never be stored in plaintext.

**Figure 3.7 Constraints/Assumptions associated with cryptographic algorithms & protocols**

Like the generation and use of random numbers, these algorithms and protocols are one of the least understood aspects in the design and implementation of a software system. It is a non-trivial task even to use them correctly. Figure 3.7 presents the constraints and assumptions associated with cryptographic algorithms and protocols. None of the constraints and assumptions associated with cryptographic algorithms and protocols address specific algorithms and protocols; instead, they focus on their usage by the software process.

### 3.3.3 An Example A Cryptographic Resources Vulnerability

Item 4 in figure 3.7 is used here to provide a detailed example of a vulnerability resulting from a violable constraint imposed by cryptographic algorithms and protocols. The vulnerable state occurs when a software process running on a computer system in a bank, sends financial transactions to a central server. Although the process encrypts all these transactions using a secret key, the format of the transactions follows an industry standard consisting of date, time, type and amount that is publicly available.

The vulnerability results from the software process using encryption to ensure both data confidentiality and integrity. A user can exploit this vulnerability by intercepting the transactions being sent to the server. Because the user does not have access to the key being used to encrypt the transactions, s/he cannot decrypt the transactions. However, because s/he is familiar with the format of the transactions, s/he can simply change the time in encrypted transactions and then resend these transactions. The consequences of the attack vary from multiple deposits to a single account to complete destruction of the transaction data held by the central server. Therefore, by failing to take this constraint into account, the software process creates a vulnerable system state.

**4.0 Conclusion and Future Work**

This report presents a taxonomy of vulnerabilities. The vulnerabilities classified by the taxonomy pertain to any software process executing above the level of the operating system. However, the taxonomy does not directly classify vulnerabilities; instead, it classifies violable constraints and assumptions. These constraints and assumptions are sources of vulnerabilities, and a vulnerability exists if they can be violated. This taxonomy is grounded in a theoretical model of computing. The model provides not only the foundation for characterizing vulnerabilities, but also the taxonomy with its classification scheme, which uses resources utilized by the software process as categories and subcategories.

We are in the process of using this taxonomy to develop a framework for deriving V&V strategies to assess software security by testing for presence of vulnerabilities. The taxonomy is a critical component of the framework, because it provides an ordered classification of known vulnerabilities. Furthermore, it affords us a distinct advantage by simplifying the process of testing a software application for presence of vulnerabilities. We can simply test whether or not the software process allows violation of a constraint or an assumption; a vulnerability exists if either can be violated.

Realizing that it is important to keep the taxonomy updated, we are working on developing a process for adding constraints and assumptions to the taxonomy. This process is important because of the dynamic nature of vulnerabilities and software security. It will ensure the longevity and usefulness of our taxonomy by keeping it updated as new vulnerabilities become known. We also recognize the importance of assessing the effectiveness of the taxonomy against real world exploits. To this end, we intend to classify vulnerabilities derived from advisories issued by CERT [11] in the year 2004. In addition to verifying the taxonomy, classifying vulnerabilities from CERT advisories will help us in evolving the taxonomy by identifying new constraints and assumptions.

The taxonomy presented in this report is distinctively different from any other existing taxonomy. It offers a completely new perspective on vulnerabilities and a novel way of classifying them. In addition to being used for assessing software security, the taxonomy can be used anywhere an ordered classification of vulnerabilities is required. Furthermore, the information embodied by the taxonomy can be used to generate requirements for creating secure software or to develop checklists for designing and implementing secure software and so forth.

**Acknowledgements**

We would like to thank Richard E. Nance, Srinidhi Varadarajan, Joe Tront, Randolph Marchany, and Sara Thorne-Thomson for their input on the early versions of the taxonomy and of this report. We would also like to thank Systems Research Center at Virginia Tech for making resources available to us to conduct this research.